\documentclass[aps,prb,showpacs,superscriptaddress]{revtex4}

\usepackage{graphicx}
\def\Vec#1{\mbox{\boldmath $#1$}}
\usepackage{dcolumn}

\begin{document}


\title{
Spectroscopy of superconducting charge qubits coupled by a Josephson inductance
}

\author{T. Yamamoto}
\affiliation{NEC Nano Electronics Research Laboratories, Tsukuba, Ibaraki 305-8501, Japan}
\affiliation{Frontier Research System, The Institute of Physical and Chemical Research (RIKEN), Wako-shi, Saitama 351-0198, Japan} 

\author{M. Watanabe}
\affiliation{Frontier Research System, The Institute of Physical and Chemical Research (RIKEN), Wako-shi, Saitama 351-0198, Japan} 

\author{J. Q. You}
\affiliation{Frontier Research System, The Institute of Physical and Chemical Research (RIKEN), Wako-shi, Saitama 351-0198, Japan} 
\affiliation{Department of Physics and Surface Physics Laboratory (National Key Laboratory), Fudan University, Shanghai 200433, China} 

\author{Yu. A. Pashkin}
\altaffiliation{on leave from Lebedev Physical Institute, Moscow 119991, Russia}
\affiliation{NEC Nano Electronics Research Laboratories, Tsukuba, Ibaraki 305-8501, Japan}
\affiliation{Frontier Research System, The Institute of Physical and Chemical Research (RIKEN), Wako-shi, Saitama 351-0198, Japan} 

\author{O. Astafiev}
\affiliation{NEC Nano Electronics Research Laboratories, Tsukuba, Ibaraki 305-8501, Japan}
\affiliation{Frontier Research System, The Institute of Physical and Chemical Research (RIKEN), Wako-shi, Saitama 351-0198, Japan} 

\author{Y. Nakamura}
\affiliation{NEC Nano Electronics Research Laboratories, Tsukuba, Ibaraki 305-8501, Japan}
\affiliation{Frontier Research System, The Institute of Physical and Chemical Research (RIKEN), Wako-shi, Saitama 351-0198, Japan} 

\author{F. Nori}
\affiliation{Frontier Research System, The Institute of Physical and Chemical Research (RIKEN), Wako-shi, Saitama 351-0198, Japan} 
\affiliation{Center for Theoretical Physics, Physics Department, Center for the Study of Complex Systems, The University of Michigan, Ann Arbor, Michigan 48109-1040, USA} 

\author{J. S. Tsai}
\affiliation{NEC Nano Electronics Research Laboratories, Tsukuba, Ibaraki 305-8501, Japan}
\affiliation{Frontier Research System, The Institute of Physical and Chemical Research (RIKEN), Wako-shi, Saitama 351-0198, Japan} 

\date{November 28, 2007}

\begin{abstract}
We have designed and experimentally implemented a circuit of inductively-coupled superconducting charge qubits, 
where a Josephson junction is used as an inductance, and the coupling between the qubits is controlled by an applied magnetic flux. 
Spectroscopic measurements on the circuit are in good agreement with theoretical calculations. 
We observed anticrossings which originate from the coupling between the qubit and the plasma mode of the Josephson junction. 
Moreover, the size of the anticrossing depends on the external magnetic flux, 
which demonstrates the controllability of the coupling. 
\end{abstract}


\maketitle

\section{\label{sec:intro}INTRODUCTION}
Recently, mesoscopic superconducting circuits have been extensively studied because 
of their potential applications for quantum information processing.~\cite{Makhlin01,Devoret03,You05,Wendin06} 
Since the first demonstration~\cite{Nakamura99} of coherent oscillations in a superconducting quantum bit (qubit), 
the quality of single qubits has increased rapidly. 
In particular, the coherence time of a single qubit has improved significantly.~\cite{Vion02} 
The mechanisms of decoherence have been intensively studied~\cite{Astafiev04,Ithier05,Bertet05,Yoshihara06,Kakuyanagi07}, 
and it is commonly accepted that in order to obtain longer coherence times, 
charge and flux qubits must be biased at the optimal point, 
where the qubit is insensitive, to first order, to fluctuations of the bias parameters. 

There has also been much progress on multiple-qubit 
systems.~\cite{Pashkin03,Berkley03,Yamamoto03,Majer05,McDermott05,Grajcar06,Steffen06,Plantenberg07}
One of the recent important topics in this field is how to best achieve controllable couplings between 
qubits,~\cite{Hime06,Ploeg07,Harris07,Niskanen07,Majer07,Sillanpaa07} 
and there are many theoretical proposals on controllable coupling 
schemes.~\cite{Makhlin99,You02,You03,Averin03,Plourde04,Lantz04,Rigetti05,Wallquist05,Brink05,Bertet06,Liu06,Niskanen06,Paraoanu06,Ashhab06,Grajcar06b,Ashhab07,Liu07,Ashhab07b} 
Although it is advantageous that two-qubit gate operations can be performed at the optimal point, 
some of the theoretical proposals cannot be used at the optimal point. 
A solution to this problem was proposed in Ref.~\onlinecite{Rigetti05}, 
who employed a technique similar to the one 
known as double-resonance in nuclear magnetic resonance.~\cite{Hartmann62} 
This and other ideas~\cite{Plourde04,Liu07} were further developed to become a so-called parametric coupling,~\cite{Bertet06,Niskanen06,Ashhab07b}
and it was recently demonstrated in time-domain experiments.~\cite{Niskanen07}

An alternative way to couple qubits at the optimal point is to use a ``longitudinal'' coupling, namely, 
an inductive coupling for charge qubits or a capacitive coupling for flux qubits.~\cite{You02,You03,Lantz04,Wallquist05,Amin05,Hutter07} 
In this case, the coupling term directly affects the energy levels at the optimal point. 

Here we report an experimental study of inductively-coupled charge qubits 
based on the theoretical study in Refs.~\onlinecite{You02} and \onlinecite{You03}. 
In this approach, an extra Josephson junction provides an inductive coupling between the qubits. 
This inductive coupling, controlled by a magnetic flux bias, is quite different from the 
usual capacitive coupling between charge qubits. 
The inductive coupling term and the single-qubit term in the Hamiltonian of the system at the charge degeneracy point 
commute with each other, which means that the eigenstates there are the tensor products of the uncoupled qubits. 
Thus, the system, when initially prepared in one of the eigenstates, does not evolve to an entangled state, 
which makes the sequence of qubit operations for the computation simpler. 
Moreover, the strength of the coupling can be controlled without changing the gate-induced charge. 
Thus, the system can always stay at the charge degeneracy point during qubit manipulations. 

Another important feature of this approach is its scalability. 
Although here we study a circuit consisting of two qubits coupled by a single Josephson junction, 
more qubits, in principle, can be coupled to the same Josephson junction, 
as discussed in Ref.~\onlinecite{You02}. 
Alternatively, one can make a one-dimensional chain of a qubit and a single Josephson junction, 
as discussed in Refs.~\onlinecite{Lantz04} and \onlinecite{Wallquist05}. 
The coupling between neighboring qubits can be controlled either by a magnetic flux bias applied to them 
or a current bias applied to the single Josephson junction between them.

\section{\label{sec:exper}EXPERIMENT}
\subsection{Circuit design}
\begin{figure}
\includegraphics[width=0.5\columnwidth,clip]{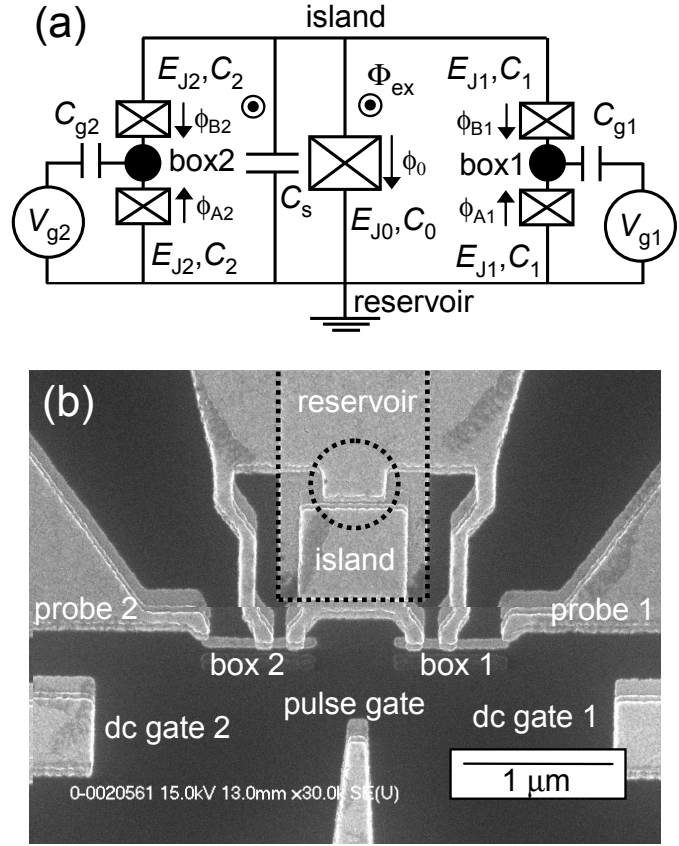}
\caption{\label{fig:fig1}
(a) Schematic circuit diagram of inductively-coupled charge qubits. 
Rectangles with an X inside denote Josephson junctions with corresponding Josephson energy $E_{Ji}$ and
junction capacitance $C_i$. 
The arrow near each junction denotes the chosen direction for the positive phase drop 
across the corresponding junction. 
(b) A scanning electron micrograph of a sample. 
A rectangular-shape electrode indicated by the dotted rectangles is connected to the reservoir 
and forms a shunt capacitance with the large island in the middle of the picture. 
The Josephson junction for the qubit coupling, inside the dotted circle, 
links the reservoir and the large island. 
Two probe electrodes and a pulse-gate electrode are shown in (b), but not in (a). 
}
\end{figure}

Figure~\ref{fig:fig1}a represents a diagram of the circuit. 
This circuit consists of two charge qubits (left and right) and a single Josephson junction 
shunted by a capacitance $C_s$ (center). 
The single Josephson junction is shared by two loops of the corresponding qubit (a split Cooper-pair box), 
where two nominally-identical Josephson junctions are attached to a superconducting small island (filled dot). 
The single Josephson junction serves as an inductor which couples the circulating currents of the two qubits. 
Because the magnitude of the circulating current depends on the flux penetrating the loop, 
we can control the strength of the coupling between the qubits by an external magnetic field. 

The Hamiltonian of this circuit, a generalization from the one for the single qubit with a large Josephson 
junction,~\cite{You06} is given by 
\begin{eqnarray}
\nonumber
H&=&\sum\limits_{i=1}^2~E_{ci}(n_i-n_{gi})^2-2E_{J1}\cos\phi_{p1}\cos\Bigl(\pi f+\frac{\phi_0}{2}\Bigr)
\\&&-2E_{J2}\cos\phi_{p2}\cos\Bigl(\pi f-\frac{\phi_0}{2}\Bigr)
+E_{c0}\Bigl(n_0+\frac{n_{g1}+n_{g2}}{2}\Bigr)^2-E_{J0}\cos\phi_0. \label{hami}
\end{eqnarray}
Here (for $i=1,2$), 
$E_{ci}=(2e)^2/4C_i$ is the Cooper-pair charging energy of the Cooper-pair box (we assume $C_{gi} \ll C_i$), 
$n_i$ is the number of excess Cooper pairs in the island, 
$n_{gi}=C_{gi} V_{gi}/2e$ is the normalized gate-induced charge on the island, 
$E_{Ji}$ is the Josephson energy of each junction of the split Cooper-pair box, 
and $\phi_{pi}$ ($=\phi_{Ai}-\phi_{Bi}$) is the total phase drop across the two junctions of the split Cooper-pair box. 
The charging energy and the Josephson energy of the coupling junction are denoted by 
$E_{c0}=(2e)^2/2[C_0+C_s+(C_1+C_2)/2]$ and $E_{J0}$, respectively. 
The phase drop across the coupling junction is denoted by $\phi_0$, and $n_0$ is its conjugate variable. 
We assume equal magnetic flux in the two loops, 
and define the relative flux bias $f=\Phi_{\rm ex}/\Phi_0$, where $\Phi_{\rm ex}$ is the magnetic flux in each loop 
and $\Phi_0$ is the flux quantum. 

When the conditions $E_{J1,2} \ll E_{J0}$ and $E_{c0} \ll E_{J0}$ are satisfied, 
the above Hamiltonian is simplified to the following effective Hamiltonian for two coupled qubits, 
based on the charge state of each Cooper-pair box,~\cite{You03} 
\begin{equation}
H_{\rm eff}=-\frac{1}{2}\sum\limits_{i=1}^2~\Bigl[E_{ci}(1-2 n_{gi})~\sigma_{zi}+2E_{Ji}^*\cos(\pi f)
~\sigma_{xi}\Bigr]+\chi~\sigma_{x1}~\sigma_{x2}. \label{effhami}
\end{equation}
Here 
\begin{equation}
\chi \equiv \chi (f)= \frac{E_{J1}E_{J2}}{4E_{J0}}\sin^2(\pi f) 
\end{equation}
is the strength of the interbit coupling, 
and 
\begin{equation} 
E_{Ji}^*=E_{Ji}\biggl[1-\frac{3}{32E_{J0}^2}(E_{Ji}^2-E_{Jj}^2)\sin^2(\pi f)\biggr], 
\end{equation} 
where $i,j=1,2~(i\neq j$). 
Note the difference in the sign of the coupling term in Eq.~(\ref{effhami}) from that in Ref.~\onlinecite{You03}. 
This is because of the different circuit geometry. 
Note also that $E_{Ji}^*$ is almost equal to $E_{Ji}$ because $E_{Ji} \ll E_{J0}$ and $E_{Ji} \sim E_{Jj}$. 
The advantage of this scheme is that the Hamiltonian [Eq.~(\ref{effhami})] at the charge degeneracy point 
$n_{g1}=n_{g2}=0.5$ consists of {\it only} $\sigma_x$ terms, 
and the eigenstates can be used as two-qubit bases because they are tensor products of uncoupled qubits. 
Moreover, we do {\it not} need to change $n_{gi}$ when we want to change the strength of the coupling, 
meaning that we can stay at the charge degeneracy point for {\it both} qubits during the qubit manipulations, 
which is preferable from the viewpoint of preserving the coherence of the qubits.~\cite{Vion02} 

Now let us consider the range of device parameters to realize this effective Hamiltonian. 
In principle, a larger $E_{J0}$ is desirable for this approach. 
However, in order to observe the effect of the coupling within the limited coherence time $T_2$, 
we cannot make $E_{J0}$ too large because the coupling coefficient $\chi$ is inversely proportional to $E_{J0}$. 
More quantitatively, it is required that 
\begin{equation}
\chi_0 \equiv \frac{E_{J1}E_{J2}}{4E_{J0}} > \frac{h}{T_2}. 
\end{equation}
Because the qubit parameters in this study are quite similar to 
those of our previous experiments,~\cite{Nakamura99,Pashkin03,Yamamoto03} 
$T_2$ is expected to be of the same order, namely $\sim$1 ns at the charge degeneracy point. 
The qubit Josephson energy $2 E_{Ji}/h$ is typically $\sim$ 10 GHz, 
meaning that $E_{J0}/E_{Ji}$ can be of order unity at maximum. 
Consequently, we designed $E_{J0}/E_{Ji}$ to be about 4 in the present study. 
Because of this limitation on the size of the coupling junction, 
the junction capacitance $C_0$ alone is not large enough to safely meet the requirement $E_{c0} \ll E_{J0}$. 
To overcome this problem, we shunted the junction by an additional capacitance $C_s$. 
However, this capacitance cannot be arbitrarily large, because we do not want to excite the 
plasma mode of the coupling junction. Therefore, here we require the condition 
$E_{Ji} \ll h\nu_p$, where $\nu_p=\sqrt{2E_{J0}E_{c0}}/h$ is 
the plasma frequency of the coupling junction. 

\subsection{Sample fabrication}
Figure~\ref{fig:fig1}b shows a scanning-electron-microscope (SEM) image of the sample. 
First, we prepared gold pads with a ground plane, 
and a coplanar waveguide on an oxidized Si substrate by a photo-lithography process. 
A 300-nm thick SiN${}_x$ grown by plasma chemical vapor deposition 
at a temperature of 250$^\circ$C was used as an insulator between the gold pads and the ground plane. 

After this photo-lithography process, the device was fabricated by a two-step electron-beam-lithography 
process using a trilayer resist 
[poly-methyl-methacrylate (PMMA)/Ge/poly-(methylmethacrylate-methacrylic acid) P(MMA-MAA), 50/20/200 nm thick]. 
In the first e-beam step, a rectangular-shape 30-nm thick Al electrode was evaporated 
after the e-beam pattern was transferred to the Ge mask and 
the bottom layer resist P(MMA-MAA) was etched by oxygen plasma. 
It is shown by the dotted rectangle in the top part of Fig.~\ref{fig:fig1}b and 
it was connected to the same gold pad (not shown) 
as the one to which the reservoir electrode was connected. 
Then the sample was brought out of the vacuum and a trilayer resist was again prepared. 
The surface of the Al electrode prepared in the first deposition 
was strongly oxidized during the following etching process (same as the one in the first step). 

Then, the coupled-qubit circuit was fabricated by a three-angle evaporation of Al (10/30/40 nm thick). 
After the evaporation of the first layer of Al, 70 mTorr oxygen was introduced into the chamber for 4 minutes typically, 
which forms tunnel barriers for the Cooper-pair boxes and the Josephson junction for the coupling (coupling junction). 
As shown by the dotted circle in Fig.~\ref{fig:fig1}b, 
the coupling junction was formed between the reservoir and the large island in the middle of the figure. 
The large island was overlapping with the rectangular-shape electrode underneath, 
forming a shunt capacitance for the coupling junction. 
From independent measurements of the current-voltage ($I$-$V$) characteristics of similar single electron transistors, 
the junction capacitance per area was estimated to be 13 fF/$\mu$m${}^2$. 
The overlapping area between the large island and the rectangular-shape electrode was 
estimated from the SEM image to be 0.71 $\times$ 0.73 $\mu$m${}^2$, 
which gives a capacitance of 6.8 fF. 

Besides those shown in Fig.~\ref{fig:fig1}a, 
there are probe electrodes for qubit readout and a pulse gate for qubit control. 
The probe electrode is attached to each box via a high-resistive tunnel junction (typically 30 M$\Omega$), 
which was formed by introducing 1 atm oxygen into the chamber for 10 minutes after the evaporation of a 
second layer of Al. 
A continuous microwave or a fast voltage pulse was applied to the pulse gate electrode, 
which is coupled almost equally to the two boxes. 

\subsection{Measurement setup}
All the measurements were performed using a dilution refrigerator at a base temperature of about 40 mK. 
DC signals were measured with a battery-powered preamplifier box. Bias voltages were supplied 
through resistive dividers and $RC$ filters in the box. For the probe bias, a voltage-feedback loop 
was also used. DC signal lines were low-pass filtered by commercial $LC$ $\pi$-filters at the top of 
the cryostat, home-made $RC$ filters at each stage of the dilution refrigerator, and the cables themselves~
(lossy CuNi coaxial cables). No DC signal lines were connected to the ground at low temperatures. 
For the transmission of high-frequency continuous microwaves or fast voltage pulses, silver-plated 
BeCu(inner)/SUS(outer) coaxial cables were used from room temperature to 4.2~K and 
Nb coaxial cables were used from 4.2~K to the base temperature. A 20-dB fixed attenuator was used at 4.2~K. 
A magnetic field was applied homogeneously to the device 
by a superconducting solenoid installed in the liquid helium bath. 

\subsection{Sample characterization} 
To characterize the device parameters, 
we first measured the $I$-$V$ characteristics of each qubit. 
We obtained the charging energy of each qubit from the slope of the Coulomb diamonds. 

\begin{figure}
\includegraphics[width=0.5\columnwidth,clip]{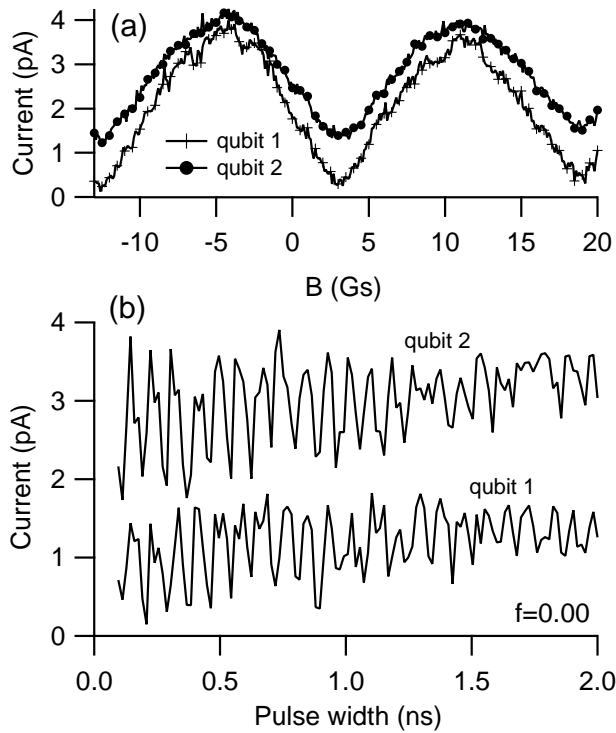}
\caption{\label{fig:fig2}
(a) Magnetic field dependence of the Josephson quasiparticle current through the two probe electrodes in sample A. 
The currents become maximum not at zero field because of the constant background field. 
(b) Coherent oscillations of sample A at the flux bias $f$=0.00. 
Two probe currents are plotted as a function of the duration time of a non-adiabatic voltage pulse 
applied to the gate electrode. For clarity, the trace for qubit 1 is offset by -0.6 pA. 
}
\end{figure}

Next we measured the field dependence of the two probe currents. 
Figure~\ref{fig:fig2}(a) shows the two probe currents of one sample (sample A) 
as a function of the external magnetic field. 
Two probes were biased at 720 $\mu$V, so that the Josephson quasiparticle (JQP) cycle~\cite{Fulton89} was activated. 
The DC gate voltages were adjusted so that each qubit was operated at the slope of the JQP peaks. 
As seen in the figure, 
the two currents are modulated by the applied magnetic field, indicating that the effective Josephson energy of 
each qubit is controllable due to the SQUID geometry. The modulation periods for the two currents are 
almost the same, as expected from nominally-equal loop sizes. 

Finally, we measured the coherent oscillations of each qubit to estimate the Josephson energy. 
Figure~\ref{fig:fig2}(b) shows coherent oscillations of the same sample at the flux bias $f$=0.00. 
Two probe currents are plotted there as a function of the duration time of a non-adiabatic voltage pulse 
applied to the gate electrode.~\cite{Nakamura99,Pashkin03,Yamamoto03} 
The measurements were done in such a way that while one qubit was oscillating at its charge degeneracy point, 
the other qubit was in the Coulomb blockade regime. 
We measured the coherent oscillations under different flux biases, 
and confirmed that the oscillation frequency showed a cosine dependence on $f$, 
from which we determined $E_{Ji}$

We fabricated samples with different $E_{Ji}/E_{ci}$ ratios of the qubit and different $E_{J0}/E_{Ji}$ ratios. 
We also fabricated a reference sample which had the same circuit geometry, 
but did not have a coupling junction. 
We summarize the parameters of the measured samples in Table~\ref{tab:parame}. 
The parameters for the coupling junction were estimated based on the junction-area 
measurements by taking scanning electron micrographs. 

\begin{table*}
\caption{\label{tab:parame}Parameters of the measured devices. 
sample E does not have a coupling junction. 
}
\begin{ruledtabular}
\begin{tabular}{cccccccccc}
 Sample & $E_{c1}/h$ & $2E_{J1}/h$ & $E_{c2}/h$ & 2$E_{J2}/h$ & $C_0$ & $C_s$ & $E_{c0}/h$ & $E_{J0}/h$ & $\nu_p$ \\ 
        & (GHz)      & (GHz)       & (GHz)      & (GHz)      & (fF)  & (fF)  & (GHz)      & (GHz)      & (GHz) \\ \hline
 A & 114 & 11.4 & 108 & 11.4 & 1.7 & 6.8 & 8.8 & 23 & 20 \\
 B & 137 & 13.6 & 127 & 12.8 & 2.0 & 6.8 & 8.5 & 36 & 25 \\
 C & 107 & 17.0 & 98 & 16.0 & 1.7 & 6.8 & 8.8 & 36 & 25 \\
 D & 111 & 23.0 & 108 & 23.0 & 1.6 & 6.8 & 8.8 & 40 & 27 \\
 E & 63.6 & 21.7 & 66.4 & 21.6 & -   & -   & -   & -    & -    \\
\end{tabular}
\end{ruledtabular}
\end{table*}

\subsection{Spectroscopic measurements}
To probe the excited states of our coupled-qubit system, 
we carried out spectroscopic measurements using a continuous microwave. 
All the samples listed in Table~\ref{tab:parame} showed qualitatively the same behavior except for sample E, 
which had no coupling junction. 
Here we focus on sample A. 

\begin{figure}
\includegraphics[width=0.5\columnwidth,clip]{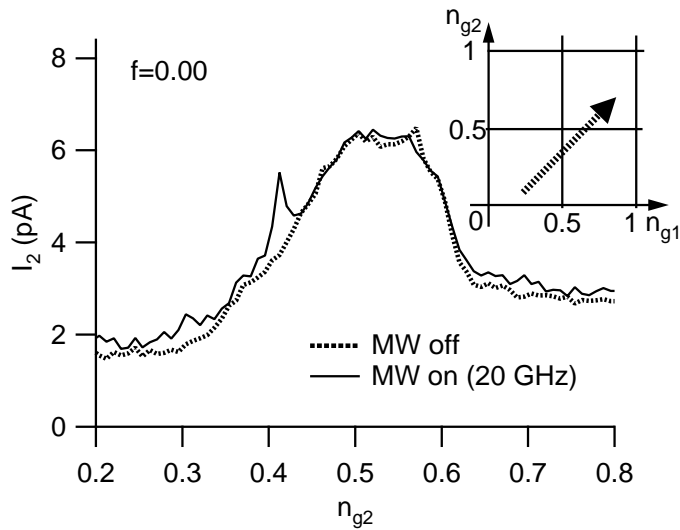}
\caption{\label{fig:fig3}
An example of the spectroscopic measurements. 
The current through probe electrode 2 is plotted as a function of $n_{g2}$, 
the normalized gate-induced charge on the gate electrode 2. 
The dotted curve is the data when no microwave is applied, while the 
solid curve is the one when a 20 GHz microwave is applied. 
The inset shows the direction of the gate sweep in the $(n_{g1}$,$n_{g2})$ plane. 
}
\end{figure}

Figure~\ref{fig:fig3} shows an example of the spectroscopic measurements. 
In the figure, the current through probe 2 ($I_2$) is plotted as a function of $n_{g2}$ 
with (solid line) and without (dotted line) microwave irradiation. 
Under microwave irradiation, besides the main JQP peak at $n_{g2}=0.50$, 
a small peak on the slope of the JQP peak is observed, which is due to the photon-assisted JQP (PAJQP) cycle.~\cite{Nakamura97} 
The peak indicates that the energy of the microwave photon matches 
the energy gap of the system at the corresponding $n_{g2}$. 
While PAJQP peaks on the left-hand side of the JQP peak correspond to a 
photon-absorption process, we could, in principle, observe PAJQP peaks on 
the right-hand side of the JQP peak as well, which correspond to a photon-emission process. 
In fact, we did observe them in some of the samples, but PAJQP peaks on the emission side were 
much weaker than those on the absorption side, as reported previously.~\cite{Nakamura97} 
In the present paper, we focus on PAJQP peaks on the absorption side. 

As shown in the inset of Fig.~\ref{fig:fig3}, 
$n_{g1}$ and $n_{g2}$ are swept simultaneously, keeping the relation $n_{g2}=\alpha n_{g1}+\beta$, 
where $\alpha$ and $\beta$ are constants. We fixed $\alpha$ to be almost equal to 1, but $\beta$ could 
vary due to background charge jumps. 
Cross capacitances, such as a capacitance between ``box 1'' and the ``dc gate 2'', are taken into account 
when we determine the relation between $n_{g1}$ and $n_{g2}$ from the relation between $V_{g1}$ and $V_{g2}$. 

In the current through probe 1 ($I_1$), we observed similar PAJQP peaks. 
Thus, we obtained the peak positions $n_{g1}^0$ and $n_{g2}^0$ at a particular microwave frequency $\nu$ 
from the $I_1$ and the $I_2$ traces, respectively. 
We repeated this measurement for different values of $\nu$ and $f$. 

\begin{figure}
\includegraphics[width=0.5\columnwidth,clip]{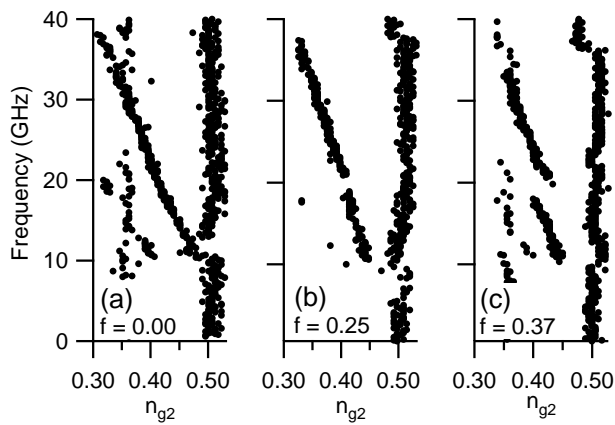}
\caption{\label{fig:fig4}
Results of the spectroscopic measurement for qubit 2 of sample A
under the flux bias (a) $f$=0.00, (b) $f$=0.25, and 
(c) $f$=0.37. The black dots represent the positions of the photon-assisted Josephson quasiparticle peaks 
at the corresponding frequencies of the applied microwave. 
}
\end{figure}

In Fig.~\ref{fig:fig4} we plot $n_{g2}^0$ as a function of $\nu$ when 
(a) $f$=0.00, (b) $f$=0.25, and (c) $f$=0.37. The step in $\nu$ is 0.1 GHz. 
The peaks observed at $n_{g2}=0.50$, independently of the microwave frequency, are the main JQP peaks. 
We sometimes observe gate-independent peaks at $n_{gi}\neq 0.50$ 
(such as those at $n_{g2} \sim 0.35$ in Fig.~\ref{fig:fig4}(a)),
which maybe due to spurious resonant modes in the surrounding circuit. 
On the left-hand side of the main JQP peaks, frequency-dependent branches are observed. 
When $f=0.00$, this branch is continuous and crosses $n_{g2}=0.50$ at 2$E_{J2}/h$. 
The overall feature looks similar to that observed in a single qubit.~\cite{Nakamura97}
When $f=0.25$, a small gap appears in the frequency-dependent branch at around 
$\nu=20$ GHz. As we increase $f$ further ($f=0.37$), the gap grows in size, becoming a clear anticrossing. 
We also observe a similar behavior in $I_1$, as shown later. 
This anticrossing is the manifestation of the {\it coupling} between the corresponding 
qubit and the plasma mode of the coupling junction. 
When $f=0.00$, the circulating current is zero, hence there is no coupling between them. 
As $f$ is increased, a circulating current develops, which gives rise to the coupling. 
This is the essence of the controllable coupling scheme in Ref.~\onlinecite{You03}. 
We note that this anticrossing was never observed at any $f$ in the sample E, 
which had {\it no} coupling junction, supporting the validity of this intuitive picture. 
In the next section, we analyze the data in a more quantitative way. 

\section{\label{sec:discus}DISCUSSION}
\subsection{Energy-band calculations}
In order to further analyze the results of the spectroscopic experiments, 
we calculated the energy spectrum of the system. 
Because Eq.~(\ref{effhami}) may not be a good approximation for our relatively small $E_{J0}/E_{Ji}$ ratio, 
we started from Eq.~(\ref{hami}). 
Here we use a method similar to the one discussed in Ref.~\onlinecite{Orlando99}. 
By considering the wave function $\Psi(\Vec{\phi})=\exp(i\Vec{k}'\cdot\Vec{\phi})\chi(\Vec{\phi})$, 
where $\Vec{k}'=[n_{g1},n_{g2},-(n_{g1}+n_{g2})]$ and $\Vec{\phi}=(\phi_{p1},\phi_{p2},\phi_0/2)$, 
we obtain a simpler Hamiltonian for $\chi(\Vec{\phi})$, 
\begin{equation}
H_0=\sum\limits_{i=1}^2-E_{ci}\frac{\partial^2}{\partial\phi_{pi}^2}-2E_{J1}\cos\phi_{p1}\cos\Bigl(\pi f+\frac{\phi_0}{2}\Bigr)
-2E_{J2}\cos\phi_{p2}\cos\Bigl(\pi~f-\frac{\phi_0}{2}\Bigr)
-E_{c0}\frac{\partial^2}{\partial\phi_0^2}-E_{J0}\cos\phi_0. \label{hami0}
\end{equation}
Here we also used the relation $n_i=-i\frac{\partial}{\partial\phi_{pi}}$ and $n_0=-i\frac{\partial}{\partial\phi_0}$.  
Because the Josephson-energy terms are periodic with respect to $\Vec{\phi}$, 
the eigenfunctions of the Hamiltonian should be of the Bloch-wave form 
\begin{equation}
\label{eq:CQ99}
\chi(\Vec{\phi})=
u\Vec{_k}(\Vec{\phi})
\,\exp(
i\Vec{k}\!\cdot\!\Vec{\phi}),  
\end{equation}
where $\Vec{k}=(k_{p1},k_{p2},2k_0)$ is the quasi-wavenumber. 
The eigenvalues of the Hamiltonian are obtained as a function of $\Vec{k}$, 
and $\Vec{k}$ is related to the normalized gate-induced charges by the periodic boundary condition $\Psi(\Vec{\phi})=\Psi(\Vec{\phi}+2\pi)$, 
namely, $n_{g1}+k_{p1}=m_1$, $n_{g2}+k_{p2}=m_2$, and $2k_0-n_{g1}-n_{g2}=2m_0$, where $m_i$'s are integers. 
We solved the central equation~\cite{KittelBook} using 1331 reciprocal lattice points.

\begin{figure}
\includegraphics[width=0.5\columnwidth,clip]{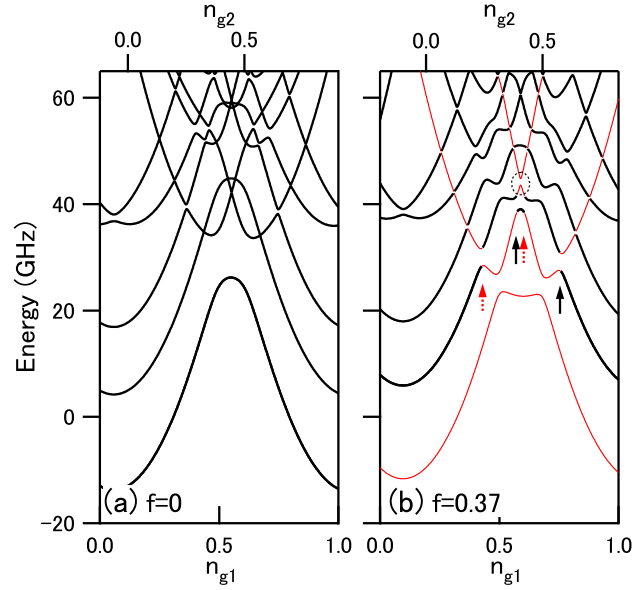}
\caption{\label{fig:fig5} (color online) 
(a) Energy spectrum of the coupled-qubit system for $f$=0 and $n_{g2}=(1.0n_{g1}-0.11)$. 
(b) Energy spectrum of the coupled-qubit system for $f$=0.37 and $n_{g2}=(1.0n_{g1}-0.21)$. 
The energy bands with zero-photon state for the plasma mode of the coupling junction are shown in thin red lines. 
Four arrows indicate the positions where anticrossings, due to the coupling between one of the two 
qubits and the plasma mode of the coupling junction, are observed. 
Two of them shown in dotted red lines correspond to the anticrossings observed in the experiment. 
A small dotted circle shows the anticrossing due to the coupling between the qubits, via the coupling junction. 
}
\end{figure}

Figure~\ref{fig:fig5} shows the energy spectrum of Hamiltonian (\ref{hami0}) 
under the flux bias (a) $f$=0.00 and (b) $f$=0.37. 
The sample parameters were taken from Table~\ref{tab:parame} (sample A). 
For the gate charges, $n_{g2}=(1.0~n_{g1}-0.11)$ was assumed for $f$=0.00, and 
$n_{g2}=(1.0~n_{g1}-0.21)$ for $f$=0.37, which are the experimental conditions used in Fig.~\ref{fig:fig4}. 

\begin{figure}
\includegraphics[width=0.5\columnwidth,clip]{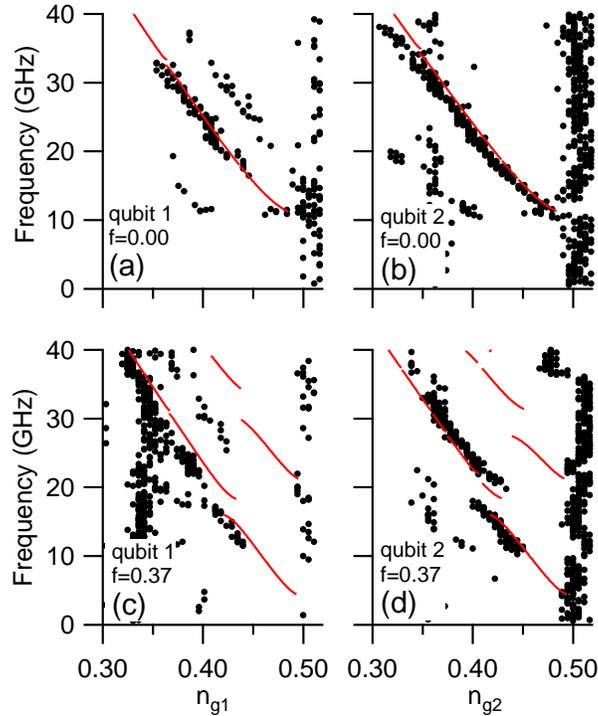}
\caption{\label{fig:fig6}
(color online) Comparison between the theory and the experimental data from the 
spectroscopic measurements. Black dots are the experimentally-obtained 
positions of photon-assisted Josephson quasiparticle peaks, and the red 
solid lines are the results of energy-band calculations. 
}
\end{figure}

Roughly speaking, the energy bands consist of those of two qubits with different 
``photon'' numbers for the oscillator of the coupling junction. 
For example, the energy bands of the two qubits with the coupling junction in the ground state 
(zero-photon state) are shown in thin red lines in Fig.~\ref{fig:fig5}(b). 
The energy gap between the ground state and the first excited state 
at the charge degeneracy point, namely, at $n_{g1}~(n_{g2})=0.5$ 
for qubit 1 (2) is equal to $2 E_{J1} \cos\pi f$ ($2 E_{J2} \cos\pi f$). 
Besides these anticrossings at the charge degeneracy point, 
there are additional anticrossings at the positions indicated by the arrows, 
where the energy bands for the zero-photon state cross with those for the one-photon state. 
Actually, the anticrossings marked by the dotted red arrows are observed in the experiment. 
The left one corresponds to the anticrossing at $n_{g1} \simeq 0.42$ in Fig.~\ref{fig:fig6}(c) 
and the right one corresponds to the anticrossing at $n_{g2} \simeq 0.42$ in Fig.~\ref{fig:fig6}(d). 
These anticrossings are the manifestation of the coupling between the corresponding qubit 
and the coupling junction. 
The anticrossings disappear at $f$=0.00, 
as seen in Fig.~\ref{fig:fig5}(a), where we expect no coupling. 

There is one more anticrossing at the point indicated by the dotted circle in Fig.~\ref{fig:fig5}(b), 
where the two energy bands with zero-photon state cross with each other. 
This anticrossing is the manifestation of the coupling between two qubits via the coupling junction, 
but its size is so small that it is hard to resolve in our spectroscopic measurements. 

Now we consider the microwave excitation from the ground state. 
In order for the excitation to a particular excited state to be observed in our readout scheme, 
that is, to be observed as an extra probe current due to a PAJQP cycle, 
the transition-matrix element between that state and the ground state must be large enough. 
In addition, there must be a large enough difference in the expectation value of the charge number for those states. 
In our calculations, we set certain thresholds for these conditions. 
In Fig.~\ref{fig:fig6}, we plot the frequency of the transitions 
which satisfy the above conditions as a function of $n_{g1}$ (left panels) 
and $n_{g2}$ (right panels), and compare with the experimental data 
(for qubit 2, the same data as in Figs.~\ref{fig:fig4}(a) and (c)). 

The overall agreement is good, considering that all the parameters used for the 
calculation are determined from {\it independent} measurements. 
As we discussed above, the anticrossings observed at $n_{g1,2} \simeq 0.42$ in Figs.~\ref{fig:fig6}(c) and (d) 
are the manifestation of the coupling between the corresponding qubit and 
the coupling junction, which can be controlled by $f$. 
In sample A, we observe split JQP peaks. 
It seems that these are two overlapping JQP peaks, which may be due to two-level charge fluctuators. 
This produces two parallel PAJQP branches as seen in Figs.~\ref{fig:fig6}(a) and (c). 

\begin{figure}
\includegraphics[width=0.5\columnwidth,clip]{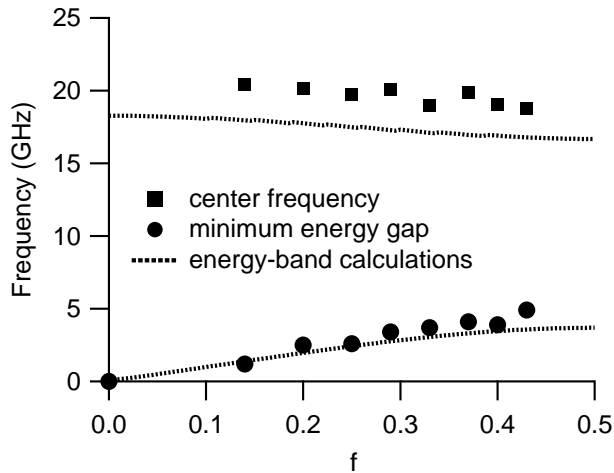}
\caption{\label{fig:fig7}
Observed ``center frequency'' and minimum energy gap of the anticrossing as a function of the reduced flux bias $f$. 
Only the data for qubit 2 is used. 
The dotted lines show the theoretical predictions from energy-band calculations. 
}
\end{figure}

We investigated the field dependence of the anticrossing further. 
From the data shown in Fig.~\ref{fig:fig4}(c), for example, 
we extracted the center frequency ($\nu_0$) and 
the minimum energy gap ($\delta \nu$) of the anticrossing. 
We analyzed the data at various $f$'s and plotted $\nu_0$ and $\delta \nu$ as a function of $f$ 
in Fig.~\ref{fig:fig7}. 
Because the data for qubit 1 was rather noisy and it was difficult to extract $\delta \nu$ and $\nu_0$, 
only the data for qubit 2 is used for this plot. 
The dotted lines are the theoretical prediction from the band calculations and 
they reproduce well the overall trend of the experimental data. 
The disagreement in the absolute value for the center frequency 
is probably due to the error in $E_{c0}$ and $E_{J0}$, which are estimated from the 
area measurements by taking scanning electron micrographs. 

Although we demonstrated the controllable coupling between the coupling junction and one of the two qubits, 
the final goal is to demonstrate the controllable coupling between the qubits. 
One demonstration would be time-domain experiments similar to those in Ref.~\onlinecite{Pashkin03}. 
When both qubits are brought to the charge degeneracy point at the same time, 
by applying a non-adiabatic voltage pulse to the gate electrode, 
the probe currents as a function of pulse width are expected to show beatings. 
In an ideal situation, namely, with a pure $|00\rangle$ as an initial state, 
with an infinitesimal rise/fall time of a non-adiabatic pulse, and with no decoherence, 
the induced probe current $I_i$ under the effective Hamiltonian (\ref{effhami}) is proportional to 
$1-\cos(2 \chi \Delta t/\hbar) \cos(2 E_{Ji} \Delta t/\hbar)$, where $i$=1, 2, and $\Delta t$ is the pulse width. 
From this formula the strength of the coupling $\chi$ can be detected as the envelope of the 
oscillations of the probe current. 
We checked in the calculation that the beating also occurs when we use the Hamitonian (\ref{hami}). 

We tried this idea in the experiments. 
We could observe the change of the oscillation frequency by applying an external magnetic field, 
but could not observe a clear change of the envelope in any of the samples listed in Table~\ref{tab:parame}. 
This is probably because of decoherence and the finite rise/fall time of the non-adiabatic voltage pulse, 
the latter of which reduces the contrast of the beatings. 
It would be interesting to test the present coupling scheme using qubits with much higher 
$E_J/E_c$ ratio, like the qubits used in Ref.~\onlinecite{Vion02}, 
with a microwave pulse instead of a non-adiabatic pulse for qubit driving.

\section{\label{sec:conlud}CONCLUSION}
We studied the spectroscopy of two charge qubits coupled by a Josephson inductance under 
various flux biases. The overall spectrum agrees well with the theory. 
We observed the anticrossings, which are the manifestation of the coupling between the coupling junction 
and one of the two qubits. The size of the anticrossing depends on the magnetic flux and disappears at zero flux, 
demonstrating that the coupling is controllable.

\section*{ACKNOWLEDGEMENTS}
The authors would like to thank Y. Makhlin, M. Wallquist, V. Shumeiko and G. Wendin for fruitful discussions, 
and S. Ashhab for valuable comments on the manuscript. 
This work was partially supported by Japan Science and Technology Agency. 
F.N. was supported in part by the U.S. National Security Agency (NSA), Army Research
Office (ARO), Laboratory of Physical Sciences (LPS), 
the National Science Foundation Grant No. EIA-0130383, and a JSPS CTC program. 
J.Q.Y. was supported in part by the National Natural Science Foundation of China
grant Nos. 10474013 and 10534060, and the NFRPC grant No. 2006CB921205.


\end{document}